\documentclass[journal=jacsat,manuscript=article]{achemso}

\usepackage[version=3]{mhchem} 
\usepackage{amssymb}

\author{Keita Funayama}
\affiliation{Toyota Central Research and Development Labratories, Inc., Nagaklute 480-1192, Japan}
\email{funayama@mosk.tytlabs.co.jp}
\author{Yuki Akura}
\affiliation{Department of Micro Engineering, Kyoto University, Kyoto 615-8540, Japan}
\author{Hiroya Tanaka}
\affiliation{Toyota Central Research and Development Labratories, Inc., Nagaklute 480-1192, Japan}
\author{Jun Hirotani}
\affiliation{Department of Micro Engineering, Kyoto University, Kyoto 615-8540, Japan}

\title[An \textsf{achemso} demo]
  {Topological Phononic Crystal on the Scale of Quasi-Ballistic Phonon Transport}

\abbreviations{IR,NMR,UV}
\keywords{American Chemical Society, \LaTeX}

\begin{document}







\begin{abstract}
Phonon engineering technology has opened up the functional thermal management of semiconductor-based classical and quantum electronics at the micro- and nanoscales. 
However, challenges have remained in designing accurate thermal characteristics based on quasi-ballistic phonon transport. 
The quasi-ballistic thermal transport arises from the combination of wave-like and diffusive phonon behaviors unlike pure diffusion. 
The topological nature has been known to be compatible with both wave and diffusive phenomena. 
Therefore, topological phononic crystals have great potential for the development of controllable and designable thermal transport based on quasi-ballistic phonons. 
In this study, we experimentally investigated the thermal behavior at the scale of quasi-ballistic phonon transport using a 1D Su-Schrieffer-Heeger model-based topological phononic crystal. 
Quasi-ballistic phonon transport was observed through change in thermal conductivity depending on the structural parameters of topological systems using micro-thermoreflectance. 
Furthermore, using topological interface states, the experimentally observed thermal behaviors were found to agree well with the theoretically expected those.
Accordingly, the topological nature is an effective approach for thermal management in micro- and nanoscale systems with quasi-ballistic phonon transport. 
Our results pave the way for a unified control scheme for wave and diffusion phenomena, such as quasi-ballistic phonons.
\end{abstract}

\section{Introduction}
Thermal management in solid nanostructures has been attracting significant attention across diverse fields such as semiconductor devices~\cite{Zhou2025,LiM2025}, microscale thermoelectric applications~\cite{Fernandez2021,Hochbaum2008,ZhangDing2022}, and thermal noise suppression and guiding for quantum devices~\cite{Ren2020,Ren2022}.
In nanoscale structures, thermal conduction is predominantly carried by phonons, which are the quantized lattice vibrations.
The thermal transport mechanism via phonons depends on the size of the structures and the mean free paths (MFPs) of the phonons.
Diffusive phonons dominate heat conduction when the size of the structures is much larger than the phonon MFP.
In contrast, ballistic phonons contribute significantly to thermal transport when the structural scale is comparable to or smaller than the MFP.
Most nanoscale solid devices consist of semiconductors; therefore, investigating phonon transport in semiconductor materials is important for the development of thermal management technologies.
Phonons exhibit ballistic behavior in many materials at low temperatures, while silicon which is widely used in practical applications has a phonon MFP Longer than one micrometer at room temperature~\cite{Anufriev2018}.
The long MFP increases the number of surface scattering events in silicon structures with characteristic lengths smaller than one micrometer, resulting in a reduction of the effective thermal conductivity~\cite{Poborchii2016,KimB2024}.
As one of promising tools to control ballistic phonon behavior, one- and two-dimensional (1D and 2D) phononic crystals (PnCs) at the sub-micron scale have been investigated~\cite{Regner2013,Anufriev2017,Verdier2017,WuY2020,Maire2017}, although diffusive phonons still remain in such small structures.
Therefore, heat is conducted by both diffusive and ballistic phonons in actual systems, which is referred to as quasi-ballistic phonon transport.

Quasi-ballistic phonons are based on quantum waves and classical diffusive phenomena.
The dual nature of waves and diffusion prevents the well-known Fourier law from accurately predicting the effective thermal conductivity and requires high computational cost to calculate it precisely using Monte Carlo simulations and first-principles calculations.

The topological nature has been reported to control both wave and diffusive phenomena.
In terms of wave phenomena, classical and quantum topological wave systems exhibit one-directional wave propagation and wave localization, with robust wave behavior against defects and perturbations~\cite{LiuY2020,MaJ2021,XueH2021,Hashemi2025,Funayama2021,Funayama2024,Isoniemi2024,Miniaci2021,TangG2024,YuS2018,HeX2019,Redon2024,ZhangQ2024,WangZ2009,ChaJ2018,Hironobu2020,Sabyasachi2018,Andrea2018}.
Thermal localization and robust decay have been reported in topological diffusion systems~\cite{HuH2022,QiM2022,WuH2023,FunayamaAPL2024,Funayama2023,LiuZ2024,Fukui2023,Yoshida2021}. 
Therefore, the combination of the topological advantages of waves and diffusion has the potential to advance thermal management technologies based on quasi-ballistic phonon transport, which exhibits the dual nature of waves and diffusion. 
However, topological heat conduction at the quasi-ballistic phonon transport scale remains largely unexplored.

Here, we experimentally demonstrate the quasi-ballistic phonon transport in 1D silicon nanostructures based on one of the most well-known topological models, the Su Schrieffer Heeger (SSH) model.
The 1D periodic structures consisted of circular sites and fine beams.
Topological interface states were excited by heating a site at the topological interface using laser irradiation.
Using micro thermoreflectance (\textmu-TR), we measured the thermal decay rate and estimated the effective thermal conductivities $k_{\mathrm{eff}}$ of the topological states.
Our fabricated structures showed a dependence of the fine beam size on $k_{\mathrm{eff}}$, indicating the contribution of ballistic phonons to thermal transport.
Further, we calculated the decay rate for the topological interface states based on the SSH model considering both diffusive and ballistic phonons that is, quasi-ballistic phonons.
For such phonon transport, the calculated decay rate is in good agreement with the experimental decay rate by using the same structural parameters and $k_{\mathrm{eff}}$.
Accordingly, these results demonstrate that topological properties can be used to engineer thermal transport via quasi-ballistic phonons.
Furthermore, topological states allow $k_{\mathrm{eff}}$ to be estimated easily by fitting the theoretical decay rate to the experimental decay rate.
We believe that our study provides a unified scheme for controlling both diffusion and wave phenomena.

\section{Results and discussion}
\subsection{Silicon-based periodic structure and experimental setups}
\begin{figure}[b!]
\centering
\includegraphics[width=1.0\textwidth]{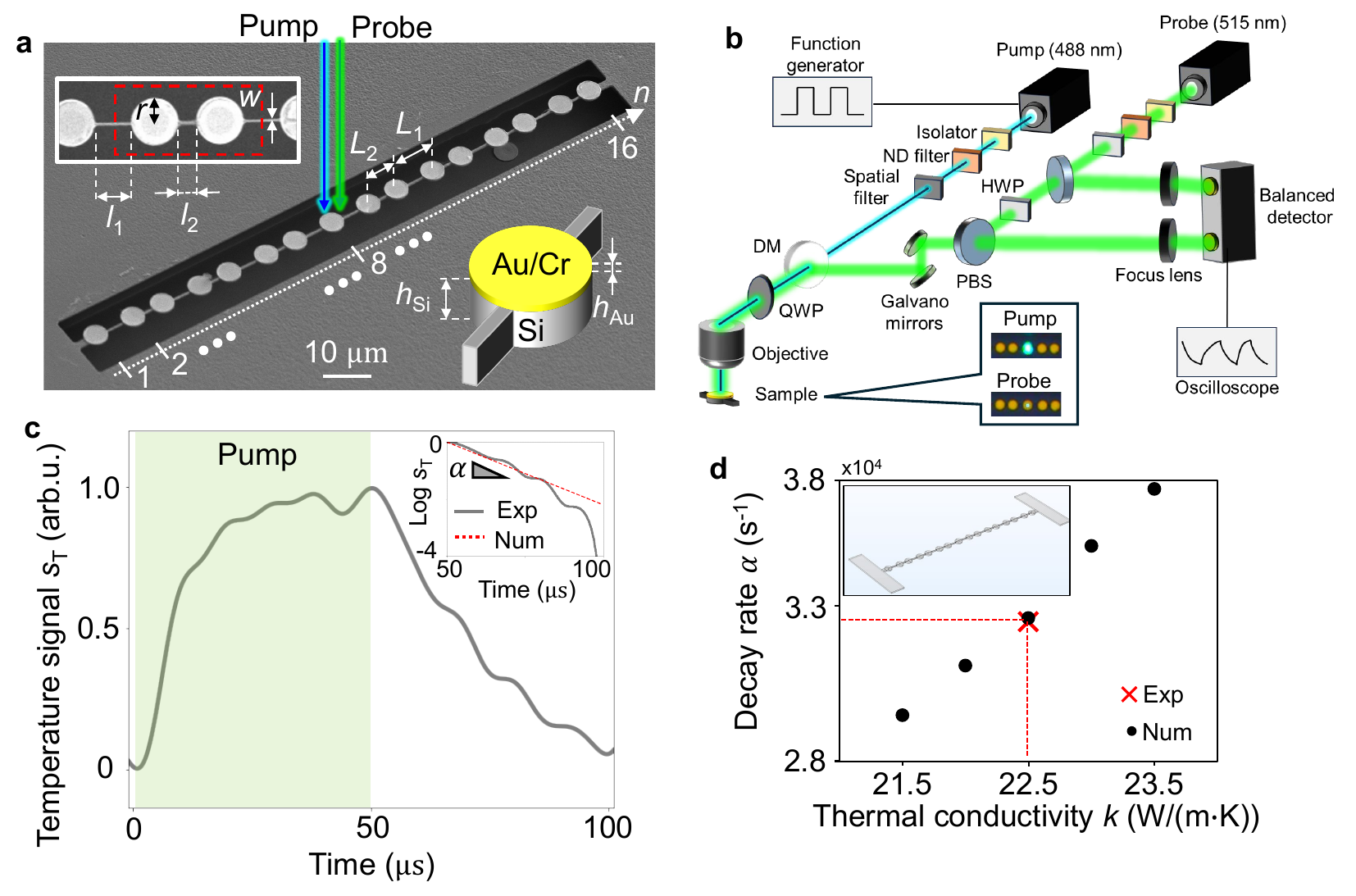}
\caption{
1D SSH model-based topological phononic crystal and experimental setup. 
(a) Scanning electron microscope (SEM) image of the suspended 1D silicon nanoscale structure consisting of 16 circular sites connected by fine beams. 
The inset of the SEM image highlights the unit cell with a red dashed line.
The schematic inset shows the structure of a circular sub-lattice.
(b) Schematic of the experimental setup for \textmu-thermoreflectance. Here, ND filter: Neutral density filter, DM: Dichroic mirror, Q(H)WP: Quarter (Half) wave plate, and PBS: Polarizing beam splitter.
(c) Time-dependent probe signal intensity $s_{\mathrm{T}}$ under 10 kHz square wave. The inset shows the logarithmic thermal decay curves between 50 and 100 \textmu s.
(d) Thermal decay rate as a function of material thermal conductivity $k$.
The effective thermal conductivity $k_{\mathrm{eff}}$ is extracted by fitting the numerical decay rate (black circles) to the experimental decay rate (red crosses).
The inset shows the numerical structure constructed to mimic the experimental device.
}
\label{fig:figure1}
\end{figure}

The silicon nanoscale structure is a 1D periodic array of unit cells consisting of two circular sub-lattices connected by fine beams, as shown in Figure~\ref{fig:figure1}a.
The details of the fabrication process are described in the Methods section.
The two circular sub-lattices, each with a radius of $r=2.5$ \textmu m, were connected alternately by fine beams of lengths of $l_1$ and $l_2$ with a width $w$, as indicated by the red square in the inset of Figure~\ref{fig:figure1}a.
Thus, the center-to-center distances between the sub-lattices are expressed as $L_{1}=2r+l_1$ and $L_{2}=2r+l_2$. 
The sub-lattices have a function as transducers from pump laser (blue line in Figure~\ref{fig:figure1}a) to heat by gold (Au) and chromium (Cr) thin films of thicknesses $h_{\mathrm{Si}}=50$ and $h_{\mathrm{Au}}=6$ nm, respectively, i.e., a total thickness is 56 nm.
In our structures, the site numbered $n=8$ in Figure~\ref{fig:figure1}a was heated by the pump laser.
We prepared nanoscale silicon structures with thicknesses of $h_{\mathrm{Si}}=0.07,~0.17,~1.0$ \textmu m to investigate thickness-dependent thermal conductivity.
In addition to thickness, the fine beams length- and width-dependent thermal conductivities were measured by using the structures with $(l_{1},~l_{2})=(10,~5),~(4,~2),~(2,~1),~(1,~0.5),~(0.4,~0.2)$ \textmu m and $w\approx0.2,~0.5,~1.0,~1.5$ \textmu m. 

The temperature of the heated sub-lattice was measured using a \textmu-TR microscope with a 515 nm probe laser (green lines in Figure~\ref{fig:figure1}a).
Figure~\ref{fig:figure1}b illustrates the experimental setup of \textmu-TR, described in detail in our previous study~\cite{Akura2025}.
A pump laser with a wavelength of 488 nm was modulated according to the pulse signal from a function generator to heat the Au/Cr film transducer. 
The reflectance of the probe laser changes with the temperature of the circular sub-lattice.
Thus, the temporal temperature on the sub-lattice was obtained from the reflected probe laser.
To reduce inherent common-mode noise in the probe laser, the reflected and reference probe laser were detected using a balanced photo detector.
Then, the output signal from the oscilloscope was then Fourier-transformed, and redundant bandwidth ranges were eliminated to improve the signal-to-noise ratio.
After applying the inverse Fourier transformation, a low-noise temporal temperature profile was obtained from the probe signal.
Before performing the thermal conduction measurements, we confirmed that each laser was accurately focused on a circular sub-lattice (Figure~\ref{fig:figure1}b).

Figure~\ref{fig:figure1}c shows the time-dependent temperature signal $s_{\mathrm{T}}$, i.e., the thermal decay curve, of site $n=8$ in the structure with $(l_{1},~l_{2})=(4,~2)$ \textmu m, $w\approx0.2$ \textmu m, and $h_{\mathrm{Si}}\approx0.17$ \textmu m.
In our experiments, the pulse laser frequency was set to 10 kHz with a 50\% duty ratio.
Therefore, the site was heated by the pump laser between 0 and 50 \textmu s (green shaded region in Figure~\ref{fig:figure1}c).
We observed that the temperature on the site exponentially decayed between 50 and 100 \textmu s.
A thermal decay rate $\alpha_\mathrm{Exp}$ is expressed as $s_{\mathrm{T}}(t)=s_{0}\mathrm{exp}(-\alpha_\mathrm{Exp} t)$, where $s_{0}$ is the maximum temperature signal value in the decay range.
Thus, we could extract $\alpha_\mathrm{Exp}$ by numerically fitting $\log(s_{\mathrm{T}}(t))$ (red dashed line) to the experimental results (gray solid line), as shown in the inset of Figure~\ref{fig:figure1}c.
$\alpha_\mathrm{Exp}$ is an average value obtained from six measurements in each structure.
Using COMSOL Multiphysics, the material thermal conductivity $k$ was treated as a free parameter to fit the numerical $\alpha$ ($\alpha_\mathrm{Num}$) to $\alpha_\mathrm{Exp}$, allowing the determination of the effective thermal conductivity $k_{\mathrm{eff}}$ of the silicon structure~\cite{Anufriev2018,KimB2024}.
Figure~\ref{fig:figure1}d shows $\alpha_\mathrm{Num}$as a function of $k$ (black circles) calculated from the structure model (inset of Figure~\ref{fig:figure1}d) similar to the experimental device.
For the numerical simulations, we investigated the temporal temperature on site of $n=8$ having initial temperature of $T_{\mathrm{in}}=398.15$ K.
The remaining structures were given an initial temperature of $T_0=298.15$ K.
More details of the numerical simulation are provided in the Methods section.
In the case shown in Figure~\ref{fig:figure1}d, the numerical result was good agreement with the experimental result (red cross) for $k=22.5~\mathrm{W/(m\cdot K)}$, therefore, we determined $k_{\mathrm{eff}}=22.5~\mathrm{W/(m\cdot K)}$.
All measurements were performed at room temperature under atmospheric pressure.
We accounted for the effects of thermal transport by radiation and air convection, and confirmed that thermal conduction through the silicon structures was dominant (details in Section 1 of supporting information).
In nanoscale structures, the heat transfer coefficient via convection has been enhanced to $\sim10^3$ W/(m$^2$ $\cdot$ K) order, and thermal emissivity has been reported to be size-dependent~\cite{XuS2024,Pulavarthy2014,Sheila2013,Ravindra2001}.
However, even under these conditions, we found that thermal conduction through the fine beams remains the dominant heat transfer path.

\subsection{Structure Size-Dependence of the Effective Thermal Conductivity}
\begin{figure}[t!]
\centering
\includegraphics[width=1.0\textwidth]{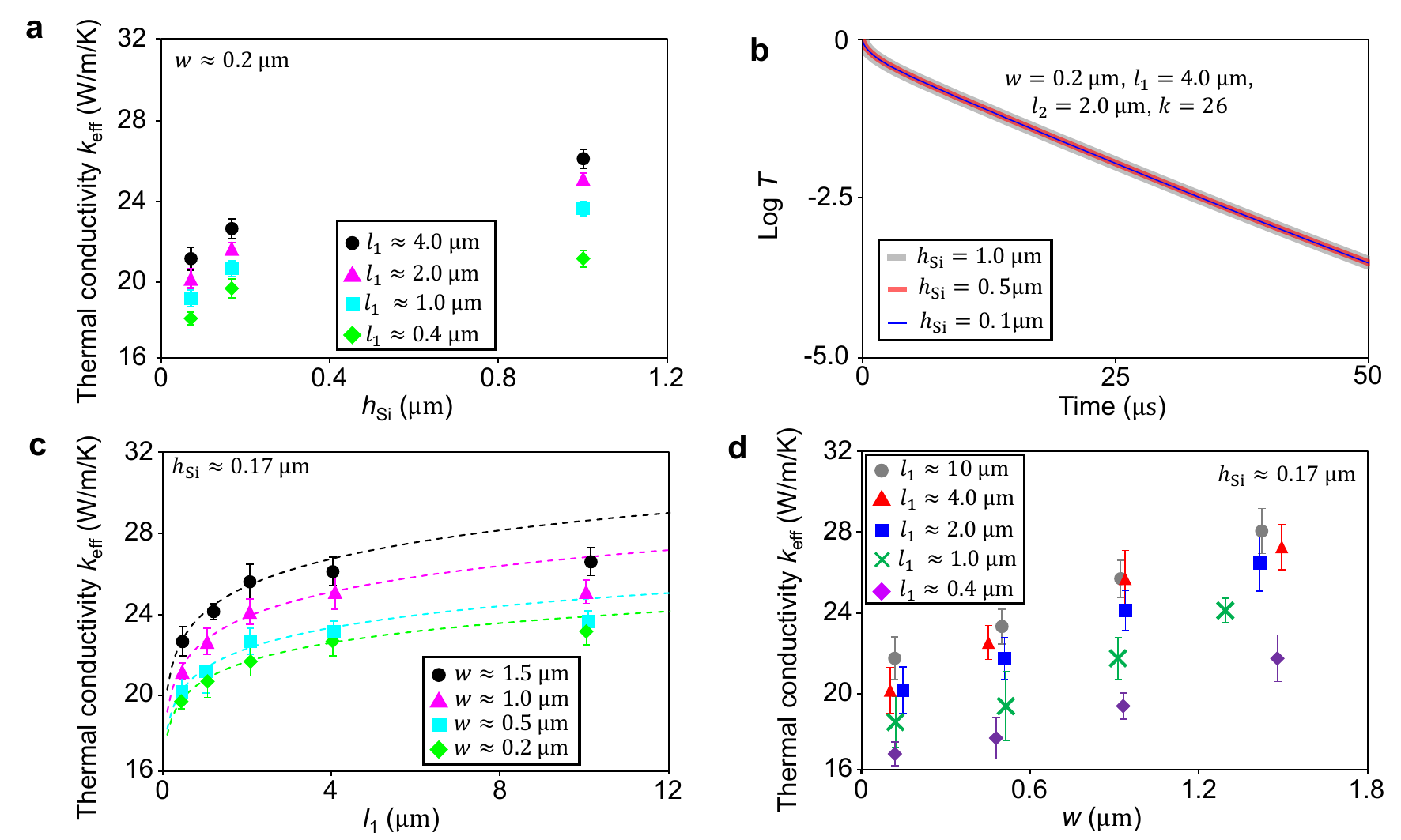}
\caption{
Size-dependent effective thermal conductivity of the 1D-silicon nanoscale structure.
(a) Effective thermal conductivity $k_{\mathrm{eff}}$ as a function of thickness of the silicon layer $h_{\mathrm{Si}}$.
Width of the fine beams $w$ is fixed at approximately 0.2 \textmu m.
(b) Logarithmic time-dependent temperatures in the numerical models with three different thicknesses $h_{\mathrm{Si}}=1.0,~0.5,~0.1$ \textmu m. The structural parameters other than $h_{\mathrm{Si}}$ are fixed.
(c) Effective thermal conductivity $k_{\mathrm{eff}}$ as a function of length of long fine beams $l_1$.
The dashed lines are proportional to ~$l_1^R$ and fitted to the experimental results with each $w$.
(d) Effective thermal conductivity $k_{\mathrm{eff}}$ as a function of width of the fine beams $w$.
In both (c) and (d), the thickness of the silicon structure $h_{\mathrm{Si}}$ is fixed at approximately 0.17 \textmu m.
}
\label{fig:figure2}
\end{figure}

We investigated the effect of the structural parameters, i.e., $h_{\mathrm{Si}}$, $l_{1\left(2\right)}$, and $w$ on $k_{\mathrm{eff}}$ to clarify the contribution of quasi-ballistic phonons to thermal transport in silicon structures.
Figure~\ref{fig:figure2}a shows the dependence of $h_{\mathrm{Si}}$ on $k_{\mathrm{eff}}$ for $w\approx0.2$ \textmu m.
We prepared the silicon layers with 1.0, 0.17, and 0.07 \textmu m thicknesses. 
Confirmation of the silicon layers thickness is in section 2 of supporting information.
We saw that $k_{\mathrm{eff}}$ decreased with decreasing $h_{\mathrm{Si}}$ regardless of the fine beam length.
The thickness dependence of $k_{\mathrm{eff}}$ is clear evidence of quasi-ballistic phonon transport.
Such a decrease in $k_{\mathrm{eff}}$ indicates that ballistic phonons scatter at the top and bottom surfaces of the fine beams.
On the other hand, when $h_{\mathrm{Si}}$ is sufficiently large, phonon scattering in the bulk, e.g., umklapp and defect scattering, becomes dominant~\cite{Gurunathan2020,JoI2013}.
In such large structures, the effects of surface scattering on thermal conduction are negligible. 
Accordingly, the material thermal conductivity $k$ in Fourier-law-based pure diffusion has no dependence on thickness of the structures.
Indeed, finite element method (FEM) based numerical results indicate identical decay curves regardless of $h_{\mathrm{Si}}$ as shown in Figure~\ref{fig:figure2}b, because FEM does not account for quantum phenomena such as ballistic phonon behavior.
Since all structural parameters other than $h_{\mathrm{Si}}$ were fixed in the numerical studies, we concluded that material thermal conductivity $k$ was the only variable parameter determining $k_{\mathrm{eff}}$ by comparing the numerical results with the experimental data in Figure~\ref{fig:figure2}a.
As discussed above, the variation in $k_{\mathrm{eff}}$ originates from ballistic phonon behavior in structures smaller than the phonon MFP.
Therefore, both ballistic and diffusive phonons contribute to thermal transport in the silicon structure.

Figure~\ref{fig:figure2}c shows $k_{\mathrm{eff}}$ as a function of $l_{1}$ for $h_{\mathrm{Si}}=0.17$ \textmu m.
The Effective thermal conductivity $k_{\mathrm{eff}}$ decreases with decreasing fine beam length $l_{1}$.
The decrease in $k_{\mathrm{eff}}$ indicates the quasi-ballistic phonon-based thermal transport along the longitudinal direction of the fine beams.
The quasi-ballistic phonon-based $k_{\mathrm{eff}}$ is proportional to $\sim {l_1}^R~(0 < R <  1)$, where $R$ is a scaling factor of ballistic phonon~\cite{Anufriev2017,Maire2017}.
The thermal transport becomes fully diffusive and fully ballistic for $R=0$ and $1$, respectively.
In our results, the nanoscale structures with each $w$ showed $R$ from 0.057 to 0.069.
Note that the fitting lines were obtained using experimental results with $l_1\leqq 2$ \textmu m, because the experimental $k_{\mathrm{eff}}$ saturates for $l_1 > 2$ \textmu m, indicating that $l_1$ exceeds the MFP of ballistic phonons.

Finally, we investigated the dependence of $k_{\mathrm{eff}}$ on $w$ as shown in Figure~\ref{fig:figure2}d. 
We found that $k_{\mathrm{eff}}$ decreased with decreasing $w$ for the structures with $l_1$.
As $w$ decreases, the side-surface scattering event of the ballistic phonons increases and prevents the ballistic phonon from functioning as a thermal transport carrier.
Based on comprehensive measurements of the size-dependent $k_{\mathrm{eff}}$, we verified the contribution of both diffusive and  ballistic phonons to thermal transport through the silicon structures.

\subsection{Topological Scheme for Quasi-ballistic Phonon Transport}
\begin{figure}[t!]
\centering
\includegraphics[width=1.0\textwidth]{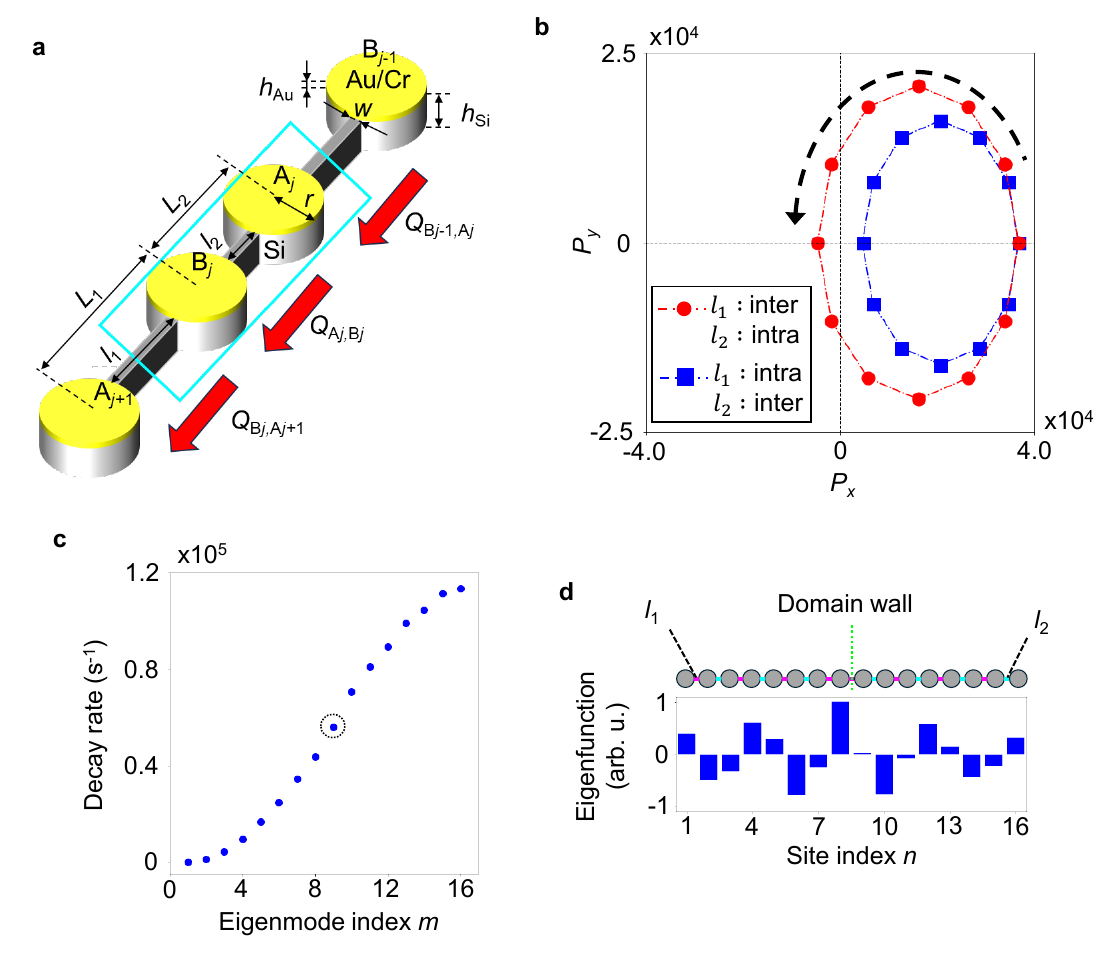}
\caption{
Topological phase transition and interface state.
(a) Schematic of thermal conduction in a 1D topological structure.
(b) Trajectories of the endpoints of the pseudospin vector $\mathbf{P}=P_\mathrm{x}\mathbf{x}+P_\mathrm{y}\mathbf{y}$. 
The winding number corresponds to the number of loops formed by the trajectories of $\mathbf{P}$.
The arrow represents the direction of the evolution of $\mathbf{P}$.
Red circles and blue squares indicate the trajectories of the topological and trivial unit cells, respectively.
(c) Retrieved spectrum of decay rates obtained by solving the eigen-equation with the Hamiltonian in Eq.~\eqref{eq:2} for $l_1=4$, $l_2=2$, $w=0.2$, $r=5$, and $h_{\mathrm{Si}}=0.17$ \textmu m.
(d) Theoretical mode profile for the topological interface state ($m=9$ in (c)) in the 1D nanoscale SSH model-based supercell.
The domain wall is located between sites $n=8$ and $9$.
}
\label{fig:figure3}
\end{figure}

From here, we analyze the quasi-ballistic phonon-based thermal transfer in terms of topological nature.
Our nanoscale silicon structure consists of an array of unit cells with two sub-lattices, A$_j$ and B$_j$, as shown in Figure~\ref{fig:figure3}a, where $j$ is the unit number.
As preparation for theoretical modeling, we define heat flow from and to site A$_{j}$ (B$_{j}$) as $Q_{\mathrm{A}_{j},\mathrm{B}_{j}}$ and $Q_{\mathrm{B}_{j-1},\mathrm{A}_{j}}$ ($Q_{\mathrm{B}_{j},\mathrm{A}_{j+1}}$ and $Q_{\mathrm{A}_{j},\mathrm{B}_{j}}$) respectively.
For $Q_{\mathrm{A}_{j},\mathrm{B}_{j}}$, the heat flow is expressed as $Q_{\mathrm{A}_{j},\mathrm{B}_{j}}=-k_{\mathrm{eff}}S(T_{\mathrm{B}_j}-T_{\mathrm{A}_j})/L_2$, where $S$ is the cross-sectional area of the beam, i.e., $S=wh_{\mathrm{Si}}$. 
Based on the heat flow and infinite periodicity with Bloch's theorem, we consider the temperature time derivatives of the temperatures $T_{\mathrm{A_j}}$ and $T_{\mathrm{B_j}}$ at each unit cell as follows:
\begin{align}
\label{eq:1}
\frac{\partial}{\partial t}
\begin{bmatrix}
T_{\mathrm{A}_j}\\
T_{\mathrm{B}_j}
\end{bmatrix}
=-
\begin{bmatrix}
D_1+D_2 & -(D_1e^{-i\beta}+D_2)\\
-(D_1e^{i\beta}+D_2) & D_1+D_2
\end{bmatrix}
\begin{bmatrix}
T_{\mathrm{A}_j}\\
T_{\mathrm{B}_j}
\end{bmatrix}
.
\end{align}
Here, $\beta$ is the wave number vector, and $D_{1(2)}$ is the effective thermal diffusivity between B$_{j-1}$ and A$_{j}$ (A$_{j}$ and B$_{j}$), i.e., $D_{1(2)}=k_{\mathrm{eff}}w/c\rho\pi r^2 L_{1(2)}$, where $c~(=700~\mathrm{J/(kg \cdot K))}$ and $\rho~(=2329~\mathrm{kg/m^3)}$ are the heat capacity and mass density of crystalline silicon, respectively (see the Methods section for details of the derivation).
While Eq.~\eqref{eq:1} is based on diffusive phonons, the effects of ballistic phonons can modify $k_{\mathrm{eff}}$, i.e., $k_{\mathrm{eff}}=k(w, l_{1,2}, h_{\mathrm{Si}})$.
In the SSH model, the topology of the structure is designable by changing the lengths of inter- and intra-beams in unit cell.
The topological phase transition is evaluated via the trajectories of the endpoints of the pseudospin vector $\mathbf{P}=P_\mathrm{x}\mathbf{x}+P_\mathrm{y}\mathbf{y}$, where $P_\mathrm{x}$ and $P_\mathrm{y}$ are derived as $P_\mathrm{x}=D_1+D_2\cos{\theta}$ and $P_\mathrm{y}=D_2\sin{\theta}$, respectively, based on the Bloch function and Pauli matrices~\cite{HuH2022,LiL2019,Jiang2020}.
Varying $\theta$ from 0 to $2\pi$, the trajectory of endpoint $\mathbf{P}$ describes a counterclockwise ellipse with circular (square) markers when the inter- and intra-beam lengths are $l_1$ ($l_2$) and $l_2$ ($l_1$), respectively, as shown in Figure~\ref{fig:figure3}b.
The winding numbers corresponding to loops around the origin are 1 and 0 for unit cells with inter-beam length of $l_1$ (topological) and $l_2$ (trivial), respectively, resulting in the topological phase transition by change of a Zak phase from 2$\pi$ to 0.
Note that the center of the ellipse for the topological phase shifts to the region at $P_\mathrm{x}>0$ because the asymmetry between $L_{1}$ and $L_{2}$ is small due to the effect of $r$.

We prepared a supercell consisting of 8-unit cells to excite the topological interface state between the topological ($1\leq j\leq4$) and trivial ($4<j\leq8$) unit cells.
Similar to Eq.~\eqref{eq:1}, we considered the temperature-time derivative of each sub-lattice in the supercell and obtained the eigenequation $\partial {\bf{T}} / \partial t=-i\mathcal{H} {\bf{T}}$, where $\mathcal{H}$ is the following Hamiltonian:
\begin{align}
\label{eq:2}
\mathcal{H}
=-i
\begin{bmatrix}
D_1 & -D_1 & 0 & \cdots & \cdots & \cdots & \cdots & 0\\
-D_2 & D_1+D_2 & -D_1 & 0 & \cdots & \cdots & \cdots & \vdots\\
\vdots & \cdots & \ddots & \cdots & \cdots & \cdots & \cdots & \vdots\\
0 & \cdots & -D_2 & D_1+D_2 & -D_1 & \cdots & \cdots & \vdots\\
0 & \cdots & \cdots & -D_1 & 2D_1 & -D_1 & \cdots & \vdots\\
0 & \cdots & \cdots & \cdots & -D_1 & D_1+D_2 & -D_2 & \vdots\\
\vdots & \cdots & \cdots & \cdots & \cdots & \cdots & \ddots & \vdots\\
0 & \cdots & \cdots & \cdots & \cdots & \cdots & -D_2 & D_2\\
\end{bmatrix}
.
\end{align}
Here, ${\bf{T}}=\left[T_\mathrm{A_1}~T_\mathrm{B_1}\cdots T_{\mathrm{A}_j}~T_{\mathrm{B}_j}\cdots T_\mathrm{A_8}~T_\mathrm{B_8}~\right]$ represents the eigenfunction as the temperature field in the structure.
By solving the eigenequation, we obtained the bulk band diagram of the supercell as shown in Figure~\ref{fig:figure3}c.
To calculate the band diagram, we substituted $l_1=4.0$, $l_2=2.0$, $w=0.2$, $h_{\mathrm{Si}}=0.17$, $r=5.0$ \textmu m, and $k_{\mathrm{eff}}=22.5$ W/(m $\cdot$ K) into the Hamiltonian in Eq.~\eqref{eq:2}.
The bulk band diagram exhibits a decay rate gap between eigenmodes $m=8$ and $10$.
The circled eigenmode ($m=9$) within this gap corresponds to a topologically protected interfacial state.
Figure~\ref{fig:figure3}d shows the mode profile of $\bf{T}$ for eigenmode $m=9$.
We see that the mode profile is localized around the domain wall with the dominant component on-site at $n=8$, i.e., sub-lattice of $\mathrm{A}_4$.
Such localized topological states have been experimentally observed by exciting only the sub-lattice containing the dominant component of the profile, since topological interface states are robust against perturbations of both profiles and structure~\cite{HuH2022}.
Accordingly, we irradiated the pump laser at site $n=8$ to experimentally excite the topological interface state.

\begin{figure}[t!]
\centering
\includegraphics[width=1.0\textwidth]{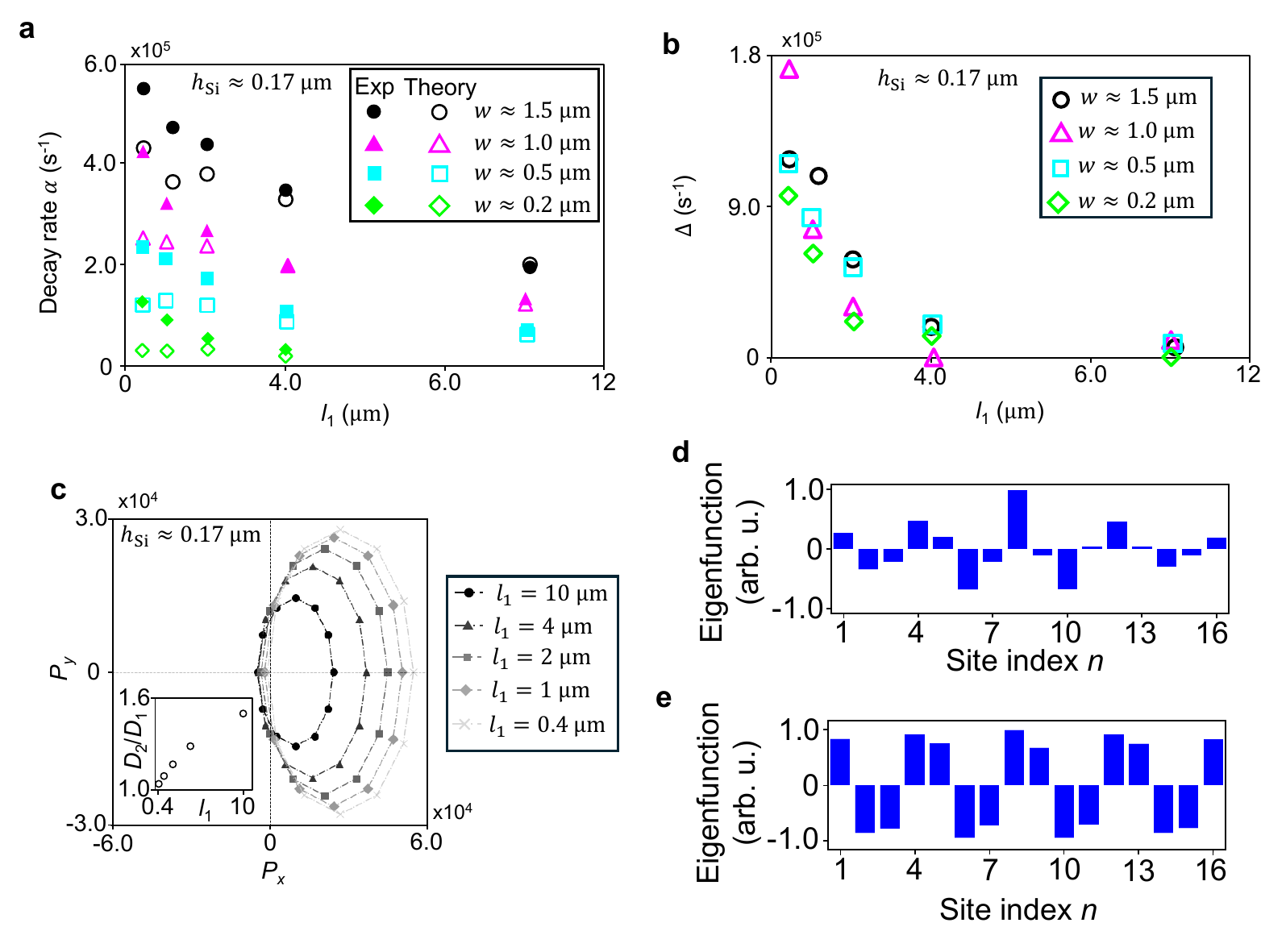}
\caption{
Comparison of experimental and theoretical decay rates for topological interface states.
(a) Decay rates $\alpha$ in the silicon nanoscale structures for topological interface states as a function of fine beam length $l_1$.
The silicon thickness $h_{\mathrm{Si}}$ is fixed at approximately 0.17 \textmu m.
The experimental results were obtained by heating and measuring the temporal temperature at site $n=8$.
(b) Discrepancy between experimental ($\alpha_{\mathrm{Exp}}$) and theoretical ($\alpha_{\mathrm{The}}$) decay rates, defined as $\it{\Delta}=\left|\alpha_{\mathrm{Exp}}-\alpha_{\mathrm{The}}\right|$, as a function of fine beam length $l_1$.
(c) Trajectories of the endpoints of the pseudospin vector $\mathbf{P}=P_\mathrm{x}\mathbf{x}+P_\mathrm{y}\mathbf{y}$ for topological unit cells with fine beam lengths $l_1=10,~4,~2,~1,~0.4$ \textmu m.
The inset shows the asymmetry ratio of the thermal diffusivity in the unit cell $D_2/D_1$ as a function of $l_1$.
Theoretical mode profiles for the topological interface states ($m=9$) for (d) $(l_1,~l_2)=(10,~5)$ \textmu m and (e) $(l_1,~l_2)=(0.4,~0.2)$ \textmu m.
}
\label{fig:figure4}
\end{figure}

The next step in this study is to verify whether the theoretical model enables to describe the quasi-ballistic phonon behavior observed in experiments.
Figure~\ref{fig:figure4}a presents the experimental decay rates (filled symbols) and theoretical decay rates (open symbols, $\alpha_\mathrm{The}$) for the nanoscale structures corresponding to the topological interface states.
Note that we calculated $\alpha_\mathrm{The}$ using the same structural parameters and $k_{\mathrm{eff}}$ as in the experimental devices.
We found that $\alpha_\mathrm{The}$ was good agreement with $\alpha_\mathrm{Exp}$ for $l_1 > 2$ \textmu m regardless of $w$.
On the other hand, for $l_1 \leq 2$ \textmu m, $\alpha_\mathrm{The}$ deviates from $\alpha_\mathrm{Exp}$.
To quantify this discrepancy, we define $\it{\Delta}=|\alpha_{\mathrm{Exp}}-\alpha_{\mathrm{The}}|$.
Figure~\ref{fig:figure4}b shows $\it{\Delta}$ as a function of $l_1$ for each $w$.
We found that $\it{\Delta}$ remains small for $l_1 > 2$ \textmu m, independent of $h_{\mathrm{Si}}$ and $w$ (section 3 of the supporting information), whereas $\it{\Delta}$ exponentially increases as $l_1$ decreases.
These results indicate that the topological scheme enables evaluation of thermal decay behavior within the quasi-ballistic phonon transport regime.
In other words, the theoretical model has a potential to identify $k_{\mathrm{eff}}$ in topologically designed structures from purely diffusive to quasi-ballistic regions by fitting $\alpha_\mathrm{The}$ to $\alpha_\mathrm{Exp}$ using $k_{\mathrm{eff}}$ as a free parameter.

We attribute the increase in $\it{\Delta}$ to the inability to experimentally excite the topological edge states by heating only site $n=8$ as the beam becomes shorter, because the localization of the mode profile is weakened.
The localization of the topological interface states becomes stronger with larger asymmetries between $D_1$ and $D_2$.
In our structures, the asymmetry between $D_1$ and $D_2$ arises solely from the difference between $L_1$ and $L_2$, since the other structural parameters are identical.
The distance between nearest-neighboring sites, $L_{1(2)}$, consists of a constant radius $r$ and a variable beam length $l_{1(2)}$.
Thus, $D_1$ and $D_2$ approach to be symmetry as $l_1$ and $l_2$ become relatively smaller with comparing with $r$.
Due to this reduction in asymmetry, Figure~\ref{fig:figure4}c shows that the trajectories of $\bf{P}$ become more horizontally aligned, causing the origin to approach the left edge of the ellipse.
While each closed ellipse still encircles the origin, the topological property weakens as the deviation between the origin and the ellipse center increases, due to chiral symmetry breaking~\cite{Asbóth2016,Zhihao2022}. 
Indeed, $D_2/D_1$ for $h_{\mathrm{Si}}\approx0.17$ \textmu m and $w\approx0.2$ \textmu m increases with beam length $l_1 (l_2)$ as shown in the inset of Figure~\ref{fig:figure4}c. 

Figures~\ref{fig:figure4}d and \ref{fig:figure4}e show the theoretical profiles of the topological interface states for supercells with $(l_1,~l_2)=(10,~5)$ and $(0.4,~0.2)$ \textmu m, respectively.
Asymmetric ratios $D_2/D_1$ were 1.5 and 1.04 for $(l_1,~l_2)=(10,~5)$ and $(0.4,~0.2)$ \textmu m, respectively, as shown in the inset of Figure~\ref{fig:figure4}c.
In Figure~\ref{fig:figure4}d, the mode profile is localized around site $n=8$, similar to that in Figure~\ref{fig:figure3}c.
Due to this localization, the topological interface state is excitable by heating site $n=8$.
By contrast, in Figure~\ref{fig:figure4}e, the mode profile is broadened, resembling the bulk modes.
For such broadened modes, we can not excite the topological interface states by heating only site $n=8$ because the eigenfunction $T_{\mathrm{A}_4}$ is not dominant in the mode profile.
Consequently, the experimentally excited mode becomes a superposition of multiple modes representing the localized profile at site $n=8$, and the observed decay rate does not correspond to that of the single topological interface state.
Indeed, $\alpha_\mathrm{Exp}$ deviates from $\alpha_\mathrm{The}$ as the fine beams shorten as shown in Figure~\ref{fig:figure4}b.
Accordingly, we can reduce $\it{\Delta}$ by decreasing the size of $r$ relative to $l_1$ and $l_2$, even for short beams.

\section{Conclusion}
We have investigated quasi-ballistic phonon transport in the SSH model-based 1D nanoscale topological structures.
We observed the size dependence of the effective thermal conductivity, $k_{\mathrm{eff}}$, to verify the contributions of both ballistic and diffusive phonons to thermal conduction.
The effective thermal conductivity $k_{\mathrm{eff}}$ decreases with decreasing thickness and width of the fine beams due to surface scattering of ballistic phonons.
The length dependence of $k_{\mathrm{eff}}$ further indicates the contribution of ballistic phonons to thermal transport through the non-zero scaling factor $R$.
We also studied quasi-ballistic thermal transport using the thermal decay rate $\alpha$ for the topological interface states in the nanoscale structures.
The experimental decay rate, $\alpha_\mathrm{Exp}$, is in good agreement with the theoretical decay rate, $\alpha_\mathrm{The}$, for topological interface states with localized mode profiles. 
This agreement demonstrates that the topological nature enables estimation of $k_{\mathrm{eff}}$ in nanoscale structures by fitting $\alpha_\mathrm{Exp}$ to $\alpha_\mathrm{The}$, using $k_{\mathrm{eff}}$ as a free parameter.
Overall, our results open up the design scheme in phonon engineering through topology.

\section{Methods}
\subsection{Numerical simulation}
We used the MEMS module in COMSOL Multiphysics 6.2 to extract the effective thermal conductivity, $k_\mathrm{eff}$, of the nanoscale structures.
To replicate the experimental conditions, we set the initial temperature $T_{\mathrm{in}}=398.75$ K at site $n=8$ in the numerical model shown in the inset of Figure~\ref{fig:figure1}d.
The initial temperature in all other regions was set to $T_0=298.15$ K.
We then simulated the temporal temperature at site $n=8$.
In the simulations, the material thermal conductivities of all regions in the silicon structure were swept to fit the numerical decay rate, $\alpha_\mathrm{Num}$, to the experimental decay rate, $\alpha_\mathrm{Exp}$.
We note that the thermal conductivity of the sub-lattices has a minimal effect on the decay behavior because the fine beams restrict the amount of heat transferred owing to their much smaller cross-section compared with that of the sub-lattices.

\subsection{Device fabrication}
We fabricated 1D nanoscale structures using a single crystal silicon-on-insulator substrate.
The thicknesses of the silicon device layer, silicon dioxide (SiO$_2$), and silicon support layer were 1, 10, and 470 \textmu m, respectively.
The silicon device layer was etched to thicknesses of 0.17 and 0.07 \textmu m using the Bosch process (MUC-21 ASE-Pegasus, Sumitomo Precision Products) to obtain device layers of different thicknesses. 
First, we patterned the circular patterns corresponding to the tranducers using electron beam (EB) lithography (JEOL JBX-6300FS) with a resist (ZEP520A, Zeon Corporation).
Then, the films of Au(50 nm)/Cr(6 nm) were fabricated by EB evaporation (Canon Anelva Corporation,
EVD-500B) to form the thermal transducers at each site.
In addition to the transducers, Au/Cr alignment patterns were simultaneously fabricated for precise alignment.
The thicknesses of the silicon layer and Au thin films were measured using a stylus profiler.
Next, we depicted patterns of the nanoscale silicon structures including the circular sub-lattices and fine beams with aligning their position to the transducers using alignment marks.
Then, the silicon device layer was etched to form the nanoscale structures.
Finally, the SiO$_2$ layer was removed by buffered hydrofluoric acid, and the silicon structures were suspended as bridge structures via supercritical drying.

\subsection{1D SSH model-based Hamiltonian}
We considered the Hamiltonian in the structure as shown in Figure~\ref{fig:figure3}a.
Assuming heat flows from site A$_j$ (B$_{j-\mathrm{1}}$) to B$_j$ (A$_j$) as $Q_{\mathrm{A}_j,\mathrm{B}_j}$ ($Q_{\mathrm{B}_{j-1},\mathrm{A}_j}$),
the temperature-time derivative of the temperature at site A$_j$ is expressed as follows: 
\begin{align}
\label{eq:3}
\frac{\partial T_{\mathrm{A}_j}}{\partial t}=-\frac{\Delta Q}{C}=-\frac{Q_{\mathrm{A}_j,\mathrm{B}_j}-Q_{\mathrm{B}_{j-1},\mathrm{A}_j}}{C},
\end{align}
where $C$ is the thermal capacity of the site.
Further elaborating $Q_{\mathrm{A}_j,\mathrm{B}_j}$ ($Q_{\mathrm{B}_{j-1},\mathrm{A}_j}$), the heat flow is rewritten as $Q_{\mathrm{A}_j,\mathrm{B}_j ({\mathrm{B}_{j-1},\mathrm{A}_j})}=-kS(T_{\mathrm{B}_j(\mathrm{A}_{j})}-T_{\mathrm{A}_j(\mathrm{B}_{j-1})})/L_{2(1)}$, where $S=wh_{\mathrm{Si}}$ is the cross section of fine beams.
Considering that $C=c\rho \pi r^2 h_{\mathrm{Si}}$, where $c$ is the specific heat, Eq.~\eqref{eq:3} is rewritten as
\begin{align}
\label{eq:4}
\frac{\partial T_{\mathrm{A}_j}}{\partial t}=\frac{kwh_{\mathrm{Si}}}{c\rho \pi r^2 h_{\mathrm{Si}}}\left(\frac{T_{\mathrm{B}_{j-1}}}{L_1}+\frac{T_{\mathrm{B}_{j}}}{L_2}-\left( \frac{1}{L_1}+\frac{1}{L_2}\right)T_{\mathrm{A}_j} \right).
\end{align}
In Eq.~\ref{eq:4}, we see that the thickness of silicon layer is canceled.
Thus, the temporal temperature does not show a thickness dependence for a constant material thermal conductivity $k$ as shown in Figure~\ref{fig:figure2}b. 
When we normalize the thermal diffusivity to $D_{1(2)}=kw/c\rho \pi r^2L_{1(2)}$ and consider Bloch's theorem, Eq.~\eqref{eq:4} is finalized as follows:
\begin{align}
\label{eq:5}
\frac{\partial T_{\mathrm{A}_j}}{\partial t}=D_1T_{\mathrm{B}_{j-1}}+D_2T_{\mathrm{B}_j}-(D_1+D_2)T_{\mathrm{A}_j}=-(D_1+D_2)T_{\mathrm{A}_j}+(D_1e^{-i\beta}+D_2)T_{\mathrm{B}_j}.
\end{align}
By deriving $\partial T_{\mathrm{B}_j}/\partial t$ similarly to $\partial T_{\mathrm{A}_j}/\partial t$, we obtain the Hamiltonian shown in Eq.~\eqref{eq:1}.

\begin{acknowledgement}
A part of this work was supported by Nagoya University microstructural characterization platform as a program of ‘Advanced Research Infrastructure for Materials and Nanotechnology in Japan (ARIM)’ of the Ministry of Education, Culture, Sports, Science and Technology (MEXT), Japan.
\end{acknowledgement}

\begin{suppinfo}

The Supporting Information is available free of charge.
Details of the estimation of the dominant thermal transport phenomenon in our experimental setup, measurement of the thickness of silicon layers, and experimental and theoretical decay rates estimated using the topological scheme for several silicon thicknesses.

\end{suppinfo}

\providecommand{\latin}[1]{#1}
\makeatletter
\providecommand{\doi}
  {\begingroup\let\do\@makeother\dospecials
  \catcode`\{=1 \catcode`\}=2 \doi@aux}
\providecommand{\doi@aux}[1]{\endgroup\texttt{#1}}
\makeatother
\providecommand*\mcitethebibliography{\thebibliography}
\csname @ifundefined\endcsname{endmcitethebibliography}  {\let\endmcitethebibliography\endthebibliography}{}

\end{document}